# A design science exploration of a visual-spatial learning system with feedback

**Kirsten Ellis**
Faculty of Information Technology
Monash University
Victoria Australia
Email: kirsten.ellis@monash.edu

**Julie Fisher**
Faculty of Information Technology
Monash University
Victoria Australia
Email: julie.fisher@monash.edu

**Louisa Willoughby**
Faculty of Arts
Monash University
Victoria Australia
Email: louisa.willoughby@monash.edu

**Jan Carlo Barca**
Faculty of Information Technology
Monash University
Victoria Australia
Email: jan.barca@monash.edu

## Abstract

Our paper is research in progress that is research investigating the use of games technology to enhance the learning of a physical skill. The Microsoft Kinect is a system designed for gaming with the capability to track the movement of users. Our research explored whether such a system could be used to provide feedback when teaching sign vocabulary. Whilst there are technologies available for teaching sign language, currently none provide feedback on the accuracy of the users' attempts at making signs. In this paper we report how the three-dimensional display capability of the technology can enhance the users' experience. Also, when using tracking to identify errors in physical movements, how and when should feedback be given. A design science approach was undertaken to find a solution to this real world problem. The design and implementation of the solution provides interesting insights into how technology cannot only emulate but also improve upon traditional learning of physical skills.

## 1   Introduction

Relatively inexpensive mainstream games technologies are now able to track users' movements in a level of detail not previously available in affordable systems. The availability of this technology enables many previously unconsidered problems to be approached but there are a plethora of assumptions that have to be made given the lack of empirical evidence on which to base design decisions. We are using a design science approach to investigate the implementation of a system to teach Australian Sign Language (Auslan) using a mainstream games technology, the Microsoft Kinect. The Kinect system uses one RGB camera and an infrared sensor to capture users' movements allowing users to interact with what is displayed on the screen. It is an affordable technology with the potential of providing an alternative to sign language classes, which are not always available particularly for those not living in a major city (Fisher et al. 2014). Australian Sign Language (Auslan) is the signed language used by the Australian Deaf community but the knowledge gained about Sign Language (SL) learning is transferable to other signed languages.

Our ongoing broader research project is seeking to answer the question "Can a mainstream game system assist people in learning Auslan?" The questions discussed in this paper are: Can the spatial capability of three-dimensional systems assist Auslan learners? How can effective feedback be provided to learners using technology to learn signs? When should this feedback be provided to the learner by the system?

We have designed and built a prototype system that is enabling us to explore possible designs for each aspect of the system. The research is an iterative process involving the design and building of a system with ongoing trialling with users to ensure a meaningful learning interaction with the system.

## 2   Background

Ninety precent of deaf children are born to hearing parents, so if they want to communicate using sign language they have the significant challenge of learning a new language (Daniels 2001; Grushkin 1998). One could argue that learning sign language is often different to learning other languages as often it is a necessity rather than a choice. For deaf people SL is the only fully accessible language so families of deaf children are forced to learn SL if they want to communicate with their child (Grushkin 1998). For SL learning, as with other physical skills, feedback to the learner from the teacher is an important component. There are methods of learning Auslan apart from classes include using videos and DVDs (Ghazarian 2005); whilst useful for learning to recognise signs these do not provide feedback on the accuracy of signs learned and do not vary the pace to meet the learner's ability (Ellis and Blashki 2004). Newer technologies have demonstrated that they can check that users are able to recognise a sign accurately but cannot provide feedback to the learner on making a sign as happens in a physical class. Unless the learner can see and understand the mistakes they might have made in making a sign the risk is the sign learned is meaningless.

Typical language learners have quite different motivations for learning a language compared with SL learners. Spoken languages are usually learned as an academic endeavour in school because they are compulsory or required for further education/ employment, or post school because the learner has an aptitude and interest in language learning. Resources for spoken languages are easy to create: for example an audio recording of native speakers using the language can be used to increase the learners' exposure to the language. Books can be used to provide supplementary resources to expose learners to the written form of the language. These are often used in tandem so that learners can hear the correct pronunciation of the words they are reading.

SLs are visual languages with no written form. Each sign consists of hand shape, location, orientation, movement and non-sign components such as facial expression (Bornstein and King 1984; Daniels 2001). Traditional attempts to teach sign language in books generally annotate line drawings or photographs of people making each sign. For example using arrows to show movement from side to side, arrows with a twist to show signs that turn or a circle or dot to show movement away from the body (Bernal and Wilson 1998; Johnston 1998). Other methods used in books to convey movement are time laps photos showing multiple frames in one picture although using this method it is difficult to interpret which is the first and last frame (Ancona and Miller 1989). Images of the side view in addition to the frontal view are sometimes used to show the maximum point that a sign moves away from the body. This can be used in conjunction with arrows to show the travel of the movement within the sign. For example, a sign cannot be accurately depicted as a picture if making the sign involves hand, finger and arm movement all of which need to be accurately executed in order for the sign to be correct. The incorporation of video into conveying signs improves the ability to convey the movement aspect of the signs although the ability to show movement away from the body is still compromised to either relying on depth perception or having to have two videos to play for a sign one which shows the front view and one which shows the side view of the sign. Creating resources for learning SL therefore is an issue compared to other languages because conveying the signs is difficult given the inherent characteristics of a SL such as Auslan.

Research has the potential to enhance the use of technology for SL learning through the development of instructional material. Ellis and Blashki (2007) developed software that incorporates vocabulary instruction, stories, songs and a game. Korte, Potter and Nielson (Korte et al. 2012; Potter et al. 2011) researched a context aware sign vocabulary learning system for young deaf children learning Auslan. Huang, Smith, Spree and Jones (2008) developed an interactive bear to teach young children using RFID flash cards. Brashear et. al. (2006) developed an American Sign Language (ASL) game CopyCat using data gloves to assist deaf children with ASL phrases. Adamo-Villani (2007) developed a Virtual Learning Environment to assist deaf children in particular aspects of Mathematics.

With the development of new technologies, research projects commenced that in addition to demonstrating SL and checking users' ability to recognize signs they also track users' ability to make signs. There are two main approaches to tracking hand movement: hardware devices and vision systems (Adamo-Villani et al. 2007). Hardware devices use direct measurement via a series of sensors to collect information such as the position of each finger and movement is collected by accelerometers that are built into data gloves (Adamo-Villani et al. 2007; Adamo-Villani and Wilbur 2007; Brashear et al. 2006; Ellis and Barca 2012).

In contrast vision systems use one or more cameras to record images of the hand then use a variety of methods to interpret the hand shape and position. Other work being undertaken by researchers involve using the Kinect sensor to recognize signs (Anjo et al. 2012; Sun et al. 2012; Zafrulla et al. 2011), however, they do not involve providing feedback to learners. The advantage of the Kinect system is that it is a single device which provides infra-red and a VGA data.

There are a few applications for mobile phones including Baby Sign and Learn (2011) and the RIDBC Auslan Tutor (2010) which are taking advantage of mobile technologies for teaching Auslan. These applications showing either video of people or character avatars demonstrating signs and they present the sign through a hierarchical text based menu system where it takes several selections in the menu to see a single sign. Auslan Signbank (Johnston 2009) provides a mobile version of a web page designed as a dictionary where a user can look up a sign from an extensive list of signs there are currently 4310 signs in the collection.

Interactions between computers and people have traditionally been limited to indirect interactions for example a keyboard or mouse. The capabilities of new games consoles have shifted expectations in the way we can interact and the type of learning possible. New games consoles are designed to be aware of the user's fine and gross motor movements with the potential to recognize and give feedback on physical skills necessary to assist with learning. More recently the new generation of technology such as touch technologies, accelerometers and motion capture devices have become mainstream particularly in mobile phones and games consoles (Peng et al. 2007). These offer great potential for teaching physical skills.

## 3    Research design

Our research approach is interpretive. Walsham (2006), argues that this is a valid approach where we are dealing with the social world and people which we are trying to make sense of. Engaged scholarship is well suited to interpretive research; it is described as "participative form of research for obtaining the advice and perspectives of key stakeholders (researchers, users, clients, sponsors, and practitioners) to understand a complex social problem." (Van De Ven 2007, p. ix).

Our research can be described as engaged scholarship, as it was undertaken by a team of academics and linguists working in close collaboration with Deaf Children Australia (DCA) and VicDeaf. These organisations have expertise in Auslan and are familiar with their communities' requirements. We used a participatory approach for all stages including sign selection and to design the interface (Fisher et al. 2014). DCA and VicDeaf were members of our advisory board where design issues were discussed. Numerous iterative cycles were also undertaken involving the deaf community to brainstorm questions such as "how would you expect this to work?" and "what would be useful here?" Users' interactions with the system were observed and to check that they understood what was presented to them and that what the system was trying to communicate was clear to the intended audience. SL classes were also observed (Willoughby et al. 2015). More details regarding how data were collected are covered later.

### 3.1    Design Science

Design science (DS) is described by (Hevner et al. 2004) as a 'problem solving paradigm' with technology as its focus and designing as the key activity. Another definition proposes that design science research is "research that develops valid general knowledge to solve field problems." (Van Aken and Romme 2009). Design science is widely recognised as an appropriate approach for conducting information systems research and has been applied widely (Gregor and Hevner 2011; March and Storey 2008; Venable et al. 2012).

March and Smith (1995) describe design science as consisting of two activities 'build and evaluate'. Building is involves constructing an artifact. Evaluating examines how well it performs. Hevner et al. (2004) describe seven guidelines for conducting design science research; these are: "Design as an artifact" requiring the development of an artifact, a relevant problem, evaluation, a research contribution, ensuring research rigour, establish the search process for the design and communicating the outcome. For our research the DS approach taken is that proposed by Venable (Venable 2006 (a); Venable 2006 (b)). Figure 1 describes our research and application of design science as described by (Venable 2006 (b)).

Venable (2006 (b)) describes a framework and context for conducting DS which is based on and extends earlier work by (Nunamaker et al. 1990/91). Venable (2006 (a); 2006 (b)) argues that this framework has a stronger focus on theory as an activity than previously published design science research (DSR). The framework proposed by Venable (2006 (b)) has four activities: 'Technology invention/design', 'Problem diagnosis', 'Theory building' and 'Technology evaluation'. Theory building is seen as central to the DSR process, "theory building is the necessary link" between the activities (Venable 2006 (b)). Next we describe each of the four activities with more details of what our research involved and the outcomes.

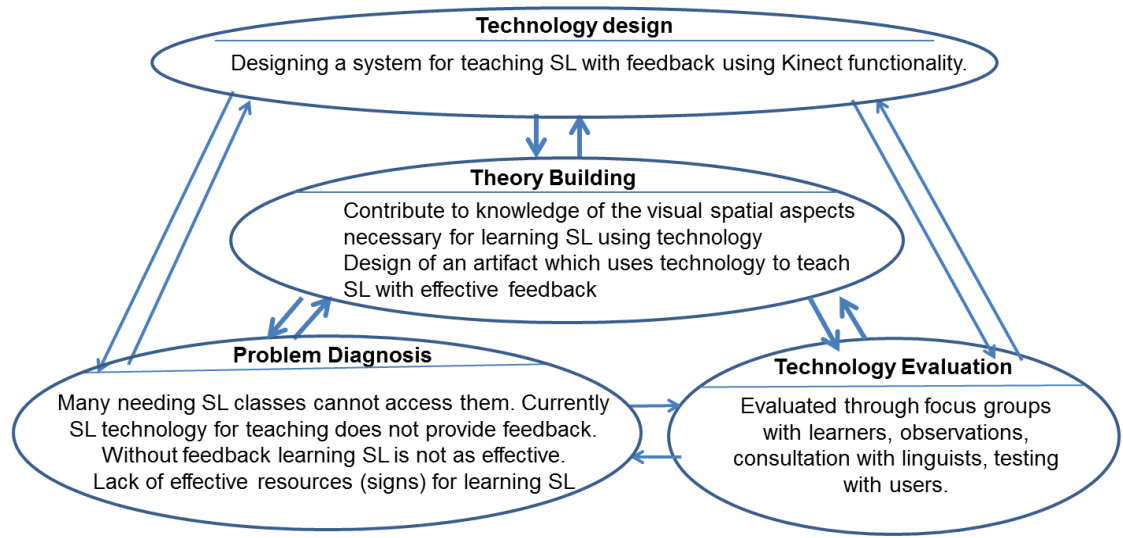

*Figure 1: Research activities*

## 3.2 Problem Diagnosis

The first step it can be argued in DS is to diagnose the problem. Venable and Baskerville (2012) suggest that the problem should be more general than specific in nature so that the resulting artefact can be applied more widely. Whilst the research context in our case was the teaching of SL, the research outcomes and the artefact have wider applicability for the teaching of other physical skills.

Hevner et al. (2004) emphasise relevance to the 'constituent community' in identifying the problem. "To be relevant to this community, research must address the problems faced and the opportunities afforded by the interaction of people, organizations, and information technology." In our case there was collaboration between researchers and practitioners in identifying the problem.

The overarching problem brought to us by DCA was how to effectively teach SL vocabulary, using technology, to people who want to learn it and cannot access classes. Further, the technology solution had to be inexpensive and something the learners could access at home. Those needing to learn Auslan, are people who are involved with those who are deaf or have other disabilities where Auslan is used for communication. The best option for learning SL is a SL class because the learner is given immediate feedback on how well they have made a given sign. However, there is a shortage of qualified Auslan teachers to give classes and in some areas, such as rural and regional areas, there are no Auslan classes available.

Designing a solution to such a problem involved understanding the component parts of the problem. These are discussed next.

### 3.2.1 Visual Spatial Requirements of Learning Sign Language

A major identified problem is the lack of effective resources for learning SL, traditional technologies for developing SL resources for teaching have not been able to deal with the visual special nature of the language; nor have they been able to provide effective feedback. It has traditionally been difficult to develop resources for teaching Auslan because of the inherent characteristics of the language; it is a visual spatial language that has no written form. As discussed earlier the visual spatial aspects of the language incorporate the elements of movement in space over time of many parts of the body. This makes conveying the language in any form where the signer is not present difficult.

### 3.2.2 Designing the type of feedback required by Auslan Learners

There is no current technology based system that provides feedback to people learning SL therefore the design of any system has to be based on how Auslan teachers provide feedback to users in Auslan classrooms. There was no prior data on the most effective feedback to provide to learners with technology. In addition, many aspects that a teacher would instinctively be able to respond to needed to be built into the system. Technology has the potential to offer alternative and more detailed feedback to learners than what is available in a teacher lead situation. Research in spoken language teaching notes that beginner students learn best when they are given explicit correction of errors: that is where the teacher highlights exactly which area of the word or sentence was produced incorrectly (Ellis et al. 2006). In order to establish appropriate forms of feedback to provide to sign language learners using technologies we conducted a study in the classroom that investigated how teachers provide feedback to students and to gain insight into there perceptions of their teaching practise. Aspects of this part of the research were recently reported in Willoughby et al. (2015).

### 3.2.3   Establishing when feedback required by Auslan Learners

Establishing the threshold of the accuracy of a sign required by learners is an important aspect to designing a technological system for learning. If the system corrected every single error then users will become frustrated as they will never be able to move on to learn the next sign especially as the error recognised by the system may have more to do with individual differences in body and movement than in the differences in the accuracy of the sign. If the system does not provide enough feedback on errors then students will assume that the signs that they are producing are correct but they may be unintelligible to the people that they will try to use the signs to communicate with. In addition to this, the level of accuracy that should be expected of a beginner would be far lower than the accuracy that would be required of a trained interpreter.

## 3.3   Technology Design

My Interactive Auslan Coach (MIAC) is the system that we developed. It uses three dimensional games technology to teach Auslan signs. Using a monitor to display a model of an avatar presenting the sign, the Microsoft Kinect is used to track the users' attempts at the sign. The users attempt is compared with the ideal model and feedback is given to the user on how accurately they have made the sign. Figure 2 is the system flowchart.

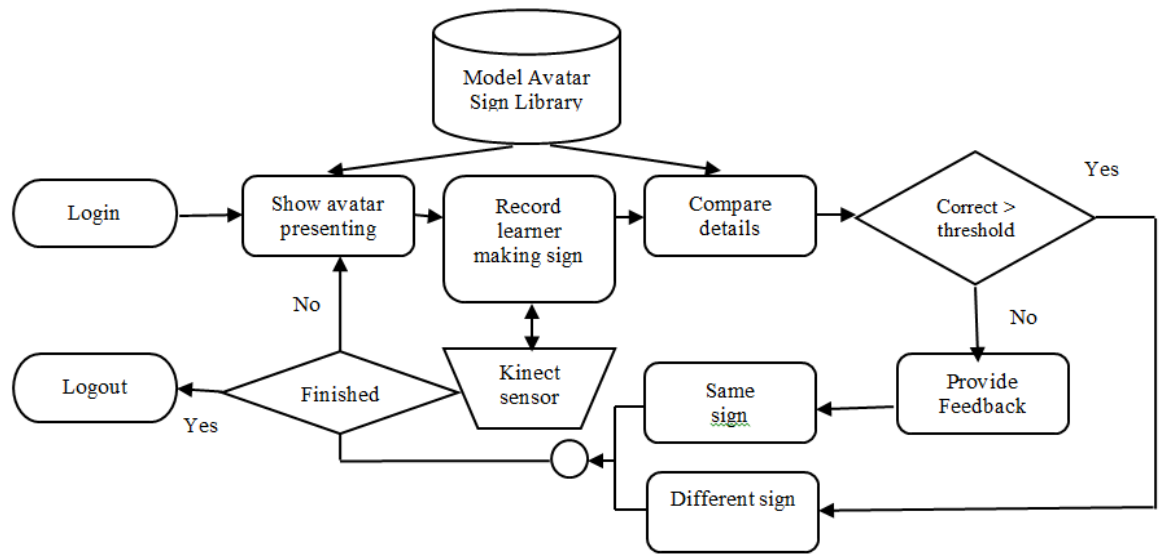

*Figure 2: Flowchart of the MIAC system*

Venable (2006 (b)) argues that the technology invention/ design is the "Enhancement or creation of a method, product, system, practice, or technique". The artifact designed through this project consisted of, as described by (Gregor and Jones 2007) the ('material artifact') insight into a two way understanding of spatial relations between signed languages and computer three dimensional games systems ('abstract artifact') based on the outcomes of the creation process.

Focus groups were conducted to provide input into our design. Participants were provided with different models of the screen including text and icons. Participants were asked to describe how they would expect the system to work based on the versions provided.

The system displays the avatar presenting a sign. A three-dimensional model of the presenter is used rather than a video of a presenter. This was an important decision as it enables different views to be shown of the presenter such as side view and close ups. In addition, the same model could be used for the presentation of the sign as for making a comparison of the sign that was recorded.

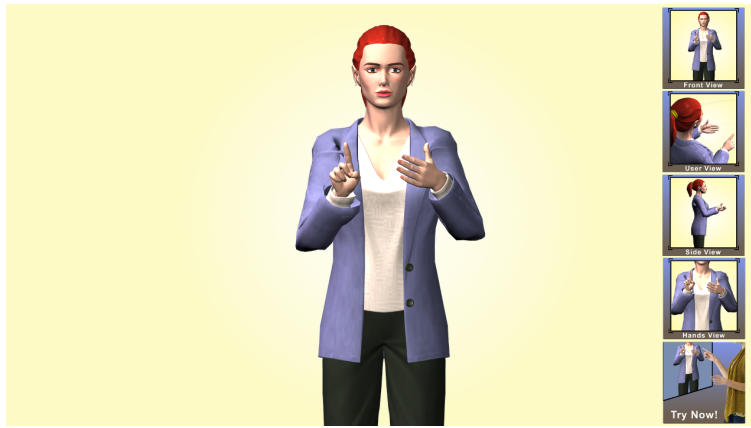

*Figure 3: Prototype of the MIAC interface*

The next phase is the recording of a learner making a sign. The Kinect sensor is used to record the movements of the learner. Because the system is not able to process the tracking at full speed a recording icon is shown on the screen during the recording phase. Consideration also had to be given to how to start and stop the recording. Also the timing had to ensure the user would get back into position after pressing the record button and then recording for a set time. This is done for convenience as the same body parts are required for turning on and off the system as is being recorded by the system.

The recording of the learner making the sign is then compared with the model signs. The comparison is made in two phases. The first phase is to check the skeleton data throughout the recording. Dynamic time warping is used to allow for variations in the timing of the sign that is presented. This means that the sign needs to have the correct movement but may have small variations in the timing of the movement and will still ensure the system recognises that the movement is correct. The second phase of the comparison is the checking of the hand shape. A swarming algorithm is used to sort through potential hand shapes to find the matching hand shape. The result of the two comparisons is an accuracy score this is compared to the acceptable threshold number which determines if the sign is considered correct or incorrect. The thresholds at the moment are not fixed until appropriate levels are determined. If the sign is incorrect then feedback is given on where the error is in the movement or the hand shape and the learner tries the same sign again. If the sign is correct the system moves on to the next sign to be learned.

### 3.3.1　Using three dimensional technology to address the visual spatial challenge

The use of three-dimensional avatars to present signs with the ability to change the viewer's perspective through changing the angle that the camera looks at the models solves one of the significant problems of developing recourses to teach signs language. The ability to move around the three-dimensional model mimics the ability of the instructor to move where they are standing in relation to the learner. This design decision was as a direct result of input from our focus group participants and Auslan learners. Using three-dimensional models to present the signs provides opportunities for displaying different perspectives. This was previously difficult to achieve using videos for presenting signs. Figures 3 is a graphic from the system; the avatar is making the sign and on the right hand side are the different viewing options users can select from.

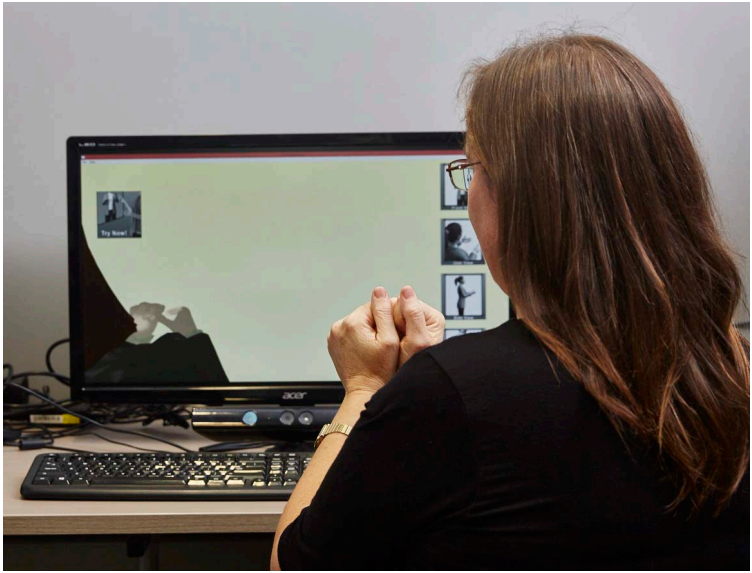

*Figure 4: A user making a sign, on the screen is the sign from over the shoulder of the user*

The ability to stand next to the instructor to see the sign from this point of view is particularly important in learning sign language because of the potential for occlusion of particular aspects of signs depending on where the learner is standing in relation to the sign that is being presented. The removal of occlusion of signs provides a solution that is even better than the traditional learning in the classroom method.

### 3.3.2 Approaches to providing feedback with technology

Given that there was no existing system for providing feedback to people learning sign language using technology, an experimental prototyping process was undertaken to ascertain a learner's perception of what different types of feedback were trying to communicate. In order to decide what type of feedback might be useful to learners we conducted an ethnographic study where we observed six teachers conducting Auslan classes and interviewed them about their teaching methods and philosophy. In line with the research on spoken language feedback Willoughby et al. (2015) also found that explicit correction was the most common correction strategy used in the beginner Auslan classroom. The reason for this is that beginner language learners often have difficulty interpreting other types of correction (such as rephrasing) as a correction, rather than say a check if the teacher has correctly heard what the student said (Lyddon 2011). As one of the teachers stated, and quoted in our paper Willoughby et al. (2015):

> *I find it effective if I explicitly highlight the source of the error and show how it is signed because if I just show the correct ones the students do not see/ realise what error they have made with the same sign*

Using the classroom feedback methods as a basis we designed a number of methods to provide feedback to the user. Although these were based on classroom methods the technology afforded alternative methods as there were no equivalents in some cases.

- Recast: The simplest feedback style is a recast, if the user doesn't get the sign correct the system repeats the sign until it is correct. This method was directly comparable to a method used in the classroom.
- Look at me look at you: Another feedback style is to provide a copy of the users attempt at the sign on screen at the same time as showing the presenter making the sign. This would be somewhat equivalent to an Auslan learner being able to see himself or herself in the mirror as they attempted to make the sign.

- Arrows to indicate path errors: to indicate where a user has made the path of a sign incorrectly an animation of their attempt at the sign is shown with arrows to indicate where their path should have gone. This would be the equivalent of a teacher pointing to where the sign should travel in space.

- Colour coding: to indicate where a user has made a sign correctly an animation of their attempt at the sign is shown, where the sign is correct the arms and hands are made in green and where there are errors in the sign the hand or limb is shown in red. There is no class teacher equivalent to this form of feedback.

- Zooming: for feedback on the errors in hand shape the system zooms in on the required hand shape so that they user can see the detail much closer. This would be the equivalent of the teacher moving closer to the student so that they can see the detail.

In the current version of the system the user can select the type of feedback that they want to receive. This is being used for the purpose of interface testing.

### 3.3.3 Using thresholds as a solution to time feedback

Sign language teachers do not correct every error that they see. In our study there were 3 teachers who corrected between 40-50% of observed errors and another 3 who corrected 75% or more of observed errors in their class. The way that teachers correct errors is often reflective on their teaching philosophy. This reflects that uncorrected errors can fossilise and correcting them immediately prevents this. However, sometimes students are not able to tell the difference between their sign and the corrected sign. One teacher recounted a time when she corrected the path of a sign for a student 10 times without the student seeing the difference. Other teachers saw mistakes as something that students would correct themselves over time especially if the student was struggling with sign formation. For example a teacher remarked (Willoughby et al. 2015)

> *It depends on whether if students are struggling or not, at times I would see that the students are struggling with their sign formation. I'd leave it at that to give them some space to learn from their own mistakes. I would suggest it to the students to watch how I sign so to learn from me and I see students correcting one's own mistakes. So this method is effective.*

Providing the correct type of feedback at the correct time is important to developing a usable system that is not frustrating. Expecting users to make perfect signs and not letting them move on to another sign would be frustrating but providing no feedback would negate the advantage of the system to improve users accuracy of the production of their signs. Currently each attempt that the user makes is given a score based on its accuracy compared to the ideal sign. For the usability testing of the system thresholds of forty, sixty and eighty precent can be selected. More research is required to determine appropriate threshold for when to give feedback to users. The acceptable thresholds will vary for user groups with different skill levels such as beginners, intermediate and advanced.

### 3.4 Technology evaluation

"Evaluation provides evidence that a new technology developed in DSR "works" or achieves the purpose for which it was designed" (Venable et al. 2012). An artifact's 'utility' should relate to IS design theory concepts or principles (Hevner et al. 2004; Venable et al. 2012). Hevner et al. (2004) and Venable (2006 (a)) identify a number of evaluation methods including observation for example of a business case study or a field study, ethnography and action research. Evaluation methods can be both qualitative and quantitative (Venable et al. 2012). Nunamaker et al. (1990/91) include theory building and observations from case and field studies when investigating systems development research projects. For our ongoing iterative research the data were drawn from a number of sources:

- Focus groups with the target audience. These were conducted with deaf and hearing people. The sessions were videoed in addition to being audio recorded so that movements could be taken into account.

- Consultations with the linguists involved in the project to consider the impact that design decisions could have on interpretation of the Auslan signs.

- Reflections from the staff at Deaf Children Australia and VicDeaf on important aspects of the design and usability of the system

- Interface testing by individuals repeatedly throughout the design and development phase. Testing was conducted with a core group of people who saw the development of the system and saw the implementation of each element of the system. In addition, new people were bought in to try the system as various stages to ensure that they could understand how to use the system without prior knowledge.

The evaluation was naturalistic and ex-post consistent with the DSR evaluation strategy proposed by Venable et al. (2012). There were many stakeholders, it was a sociotechnical artifact and the evaluation explored its effectiveness. It involved real users with a real problem and a real system. The 'evaluands' were the artefact and the method (Venable et al. 2012). The design, development and implementation of this system is ongoing.

### 3.5 Theory Building

Hevner et al. (2004) argue, a clear contribution from the research must be provided which answers the question "What are the new and interesting contributions?" This is usually an artifact. For the artifacts to be of value they must be able to be used in the context for which they were designed and to be beneficial and relevant to practitioners (Hevner et al. 2004). March and Smith (1995) propose that in IT research we should be proposing theories and explaining the characteristics of the IT artifact in its operating context and the uniqueness of the artifact. In the context of DSR, Gregor and Jones (2007) argues that "a theory can be about both the principles underlying the form of the design and also about the act of implementing the design in the real world (an intervention)." There should also be some contribution in terms of how such knowledge might be applied (Gregor and Jones 2007).

This research contributes to knowledge in the visual spatial aspects of learning sign language in the context of building technology based SL learning systems. Such research has not previously been conducted. The knowledge that sign language users have about the use of space in Auslan, see e.g. Johnston and Schembri (2007) could be useful to interface designers. This previous work informs the emergent natural user interface use of space that has previously not been considered. Auslan has an established acceptable natural signing space. Most signs are made in neutral signing spaces, which is the comfortable space in front of the user's chest. There are also some signs that are made close to the face. These can have many different, close together locations because the accuracy of the location can be judged in relation to the face, which is a much finer space. Signs are made as low as the hands resting at the side of the leg but they are never made any lower. Natural sign space allows for signs to be made above the head and very occasionally at arms reach at the side. Signs are never made behind the head or behind the back because the signs would be not be able to be seen. Our research findings in relation to the signing space used in Auslan, will inform the comfortable and acceptable space for new motion tracking games systems such as the Microsoft Kinect.

This research also revealed that technology has the potential to assist in the classroom by providing alternative viewing perspectives for people learning sign language. Further research is needed to solve the problem of occlusion in the traditional classroom setting. Classroom teachers could be supported by the use of a camera that projects a sign from their perspective to a monitor that displays the image on a monitor behind the teacher this could improve the student comprehension of details of the signs and thereby improve the traditional classroom learning experience.

## 4   Discussion

Using the designs science methodology was an approach that enabled research to be conducted that would not be possible using another research method. Once the problem was identified the system was designed and parts of the system were developed. It was the partially developed system that enabled the users to articulate aspects of the problem that were not revealed when talking in abstract terms. The discovery of the solution of providing multiple views of the signs being made did not come up until the users saw the system and made the researchers aware of the issue.

It was only in the interaction with sign language users that it became apparent to the researchers that the spatial issues in signing have a number of aspects in common with the some of the elements of the relatively new motion tracking systems. They are both concerned with where relative to the body it is comfortable to make gestures. They both also have to deal with issues of occlusion. There is a feedback loop that can take the elements of a technological system and use it to improve the traditional systems. The traditional system can then feed the knowledge gained through the interaction back into the technological system so that both are improved.

The research for this particular system is still working to find a solution to providing feedback to people learning SL but the insights gained can be applied to a whole range of physical skills. SL is an extremely complex physical skill so if technology can provide feedback for this skill then it would be able to provide feedback on a whole range of other movements. One could see application for the technology in sports, music and medical purposes such as rehabilitation of patients' movements. The low cost of the mainstream technology makes it a solution that could be implemented on a large scale, which has not been possible previously.

The research is still in progress. The next step is to conduct user evaluations on a small groups of people to ensure that user can work through the program on their own and then to conduct an evaluation on a larger sample of people to explore how the system works as a learning tool.

## 5   Conclusion

This research used a design science methodology to explore the development of a solution to the real world problem of using technology to teach sign language. Sign language is a visual spatial language with no spoken or written form and that makes it challenging to develop learning resources. The spatial nature of sign language was well matched to the three-dimensional capability of the Kinect system and it enables the learner to view a sign from multiple perspective which helped to clarify the requirement of the sign. The development of an artifact that utilized games tracking technology was a novel solution for to how to record the users attempts at signs and then to provide feedback on the accuracy of their attempts at making the signs. The frequency of feedback has a significant impact on the effectiveness and usability of a learning system so developing thresholds is important to optimizing learning outcomes. These were solutions to research questions that could only be revealed by carrying out the design science process of building and evaluation.

## Acknowledgements

We would like to thank the contribution of our partners in this research. This project was funded through an ARC Linkage Grant. Our industry partners were Deaf Children Australia and VicDeaf. Microsoft provided the Kinect technology.